\documentclass[num-refs]{wiley-article}

\usepackage{multirow,multicol}
\usepackage{graphicx}

\usepackage{siunitx}
\usepackage{tabularx}
\usepackage[normalem]{ulem}

\papertype{Research Article}
\title{A wireless bilateral transceiver coil based on volume decoupled resonators for a clinical MR mammography}

\author[1]{Pavel M. Tikhonov}
    \affil[1]{School of Physics and Engineering, ITMO University, St. Petersburg, Russian Federation}
    \author[1]{Alexander D. Fedotov}
    \affil[2]{High-Field Magnetic Resonance, Max Planck Institute for Biological Cybernetics, Tübingen, Germany}
    \author[2]{\\ Georgiy A. Solomakha} 
    \author[1]{Anna A. Hurshkainen}

\corraddress{Pavel~Tikhonov, School of Physics and Engineering, ITMO University, St. Petersburg, 197101, Russia}
\corremail{pavel.tikhonov@metalab.ifmo.ru}

\fundinginfo{This work was supported by state assignment No. FSER-2022–0010 within the framework of the national project “Science and Universities”}

\runningauthor{Tikhonov et al.}

\begin{document}
\maketitle
\begin{abstract} 
\small

Wireless radio frequency coils provide a promising solution for clinical MR applications due to several benefits, such as cable-free connection and compatibility with MR platforms of different vendors. Namely, for the purpose of clinical high-field human breast imaging several wireless transceiver coils are known to the date, those operational principle is based on inductive coupling with a body coil. These coils are commonly consist of a several volume resonators to perform bilateral breast imaging. Due to the electrically close location of volume resonators, strong inductive coupling is observed, resulting in the occurrence of hybrid modes. In principle, MR imaging using one of the hybrid modes is possible provided by the homogeneity of a $B^{+}$ distribution. However, the question of influence of volume resonators coupling on wireless coil transmit efficiency and receive sensitivity was not previously studied. By this work, we performed study to understand this issue. The first wireless coil with decoupled resonators is developed, evaluated numerically and experimentally including in vivo study on healthy volunteers. According to the obtained results, transmit efficiency and receive sensitivity of a pair of decoupled Helmholtz resonators is at least 24\% higher than for a pair of coupled resonators.

\keywords{wireless coil, electromagnetic decoupling, Helmholtz resonator, signal-to-noise ratio}

\end{abstract}


\section{Introduction}

Clinical breast MR imaging is a valuable tool for diagnostics of a breast cancer, that has significant social impact ~\cite{Breast_cancer_2005, Breast_cancer_2017}. To obtain clinically relevant human breast images, multichannel Rx-only coils are commonly used with the number of receive channels up to sixteen[vendor ref]. However, availability of dedicated breast coils in clinics is low ~\cite{Breast_cancer_2017} due to high cost, especially for coils with high number of receive channels. In this sense, wireless dedicated breast coils provide a good alternative to cable-connected Rx coils ~\cite{Puchnin_MRM_2022, Shchelokova_NC_2020}. Such wireless coils couple inductively with a body coil of MR scanner and operate as a local booster of $B_1$ in the target region of interest (ROI) of a human breast. A transceiver body coil is used in this case in both transmit and receive modes, while a local wireless coil can be used as a Tx/Rx coil ~\cite{Puchnin_MRM_2022, wu2024wireless, lu2023low, knee_meta_coil, Chi_2021, Shchelokova_MRM_2018, 2021wirelessRFarray} as well as a Rx-only coil ~\cite{stoja2021improving, saha2020smart, zhu2023helmholtz}. 

The main challenge associated with an application of wireless coils in clinical practice is their receive sensitivity. To obtain clinically relevant images with an appropriate speed, parallel imaging (PI) technique is ubiquitously used for Rx-only coils ~\cite{PI_2004}. However, operational principle of wireless coils inductively coupled with a body coil does not support a PI technique which imposes strong requirements for receive sensitivity of wireless coils. To improve it, quadrature configuration of bilateral wireless coil was proposed recently ~\cite{Puchnin_MRM_2022} using the benefits of circularly polarized body coil.

Wireless coils for human breast imaging known to the date consist of volume resonators of different configurations ~\cite{Puchnin_MRM_2022,Shchelokova_NC_2020}. To create a bilateral coil setup, at least two volume resonators are needed located electrically close to each other. This close location results in strong electromagnetic coupling and occurrence of hybrid eigenmodes. MR imaging using one of these eigenmodes is possible if its resonant frequency is tuned to the Larmor frequency of a particular MR platform. If the eigenmode near field pattern is similar in both of the bilateral configuration halves, the relevant MR image could be obtained which was shown in ~\cite{Puchnin_MRM_2022}.

It is known from the theory of multichannel MR coil techniques that decoupling of antenna elements reduces parasitic influence of inductive coupling on MR coil performance ~\cite{roemer_nmr_1990}. In particular, noise correlation is reduced, improving the performance of parallel imaging techniques. Moreover, reduction of destructive interference of the primary and secondary magnetic fields occurred in the ROI due to coils coupling improves homogeneity of $|B1^{+}|$ in the target area ~\cite{Hurshkainen_wires}.

In this work, the coil decoupling approach is applied and evaluated in terms of its influence on the receive sensitivity of a wireless coil based on a pair of Helmholtz coils. At first, two pairs of coupled and decoupled using a transformer Helmholtz resonator are evaluated using numerical simulations and on-bench test with a homogeneous phantom to calculate resonant properties of setups under the study. Moreover, electromagnetic simulations with a human voxel model were also performed to evaluate the $|B^{+}|$ and SAR distributions. Experimental tests using MRI scanners were also performed with homogeneous phantom and, finally, on healthy volunteers to evaluate the results of numerical simulations. The predicted positive influence of volume resonator decoupling on a receive sensitivity of a wireless bilateral breast coil was confirmed numerically and experimentally.

\section{Methods}
\subsection{Wireless coil design and simulations}

\begin{figure*}[ht!]
\centering
\includegraphics[width=1\linewidth]{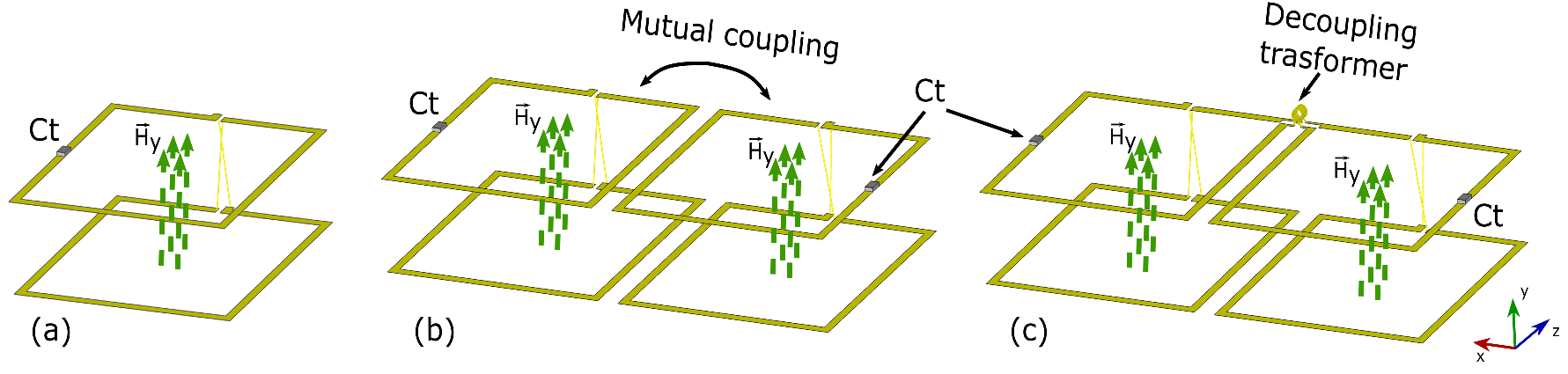}
\caption{Wireless coil configurations: a single Helmholtz coil (a), a pair of inductively coupled Helmholtz coils (b) and a pair of Helmholtz coils decoupled by a transformer (c)}
\label{Fig1}
\end{figure*}

The proposed coil consists of pair Helmholtz resonators forming a bilateral transceiver wireless coil. Helmholtz resonators (Figure \ref{Fig1}a) provide side access to human tissues in the case when biopsy procedure is required. Moreover, such resonators have simple configuration compare to analogues ~\cite{Puchnin_MRM_2022, knee_meta_coil} allowing easy tuning and decoupling. To create a bilateral coil, a pair of volume resonators is needed, as it is shown in Figure \ref{Fig1}b. Since two resonators are located electrically close to each other, mutual inductive coupling is expected to occur. To evaluate this, numerical simulations using a frequency domain and time solvers of CST Microwave Studio 2022 were performed. 

\begin{figure*}[ht!]
\centering
\includegraphics[width=1\linewidth]{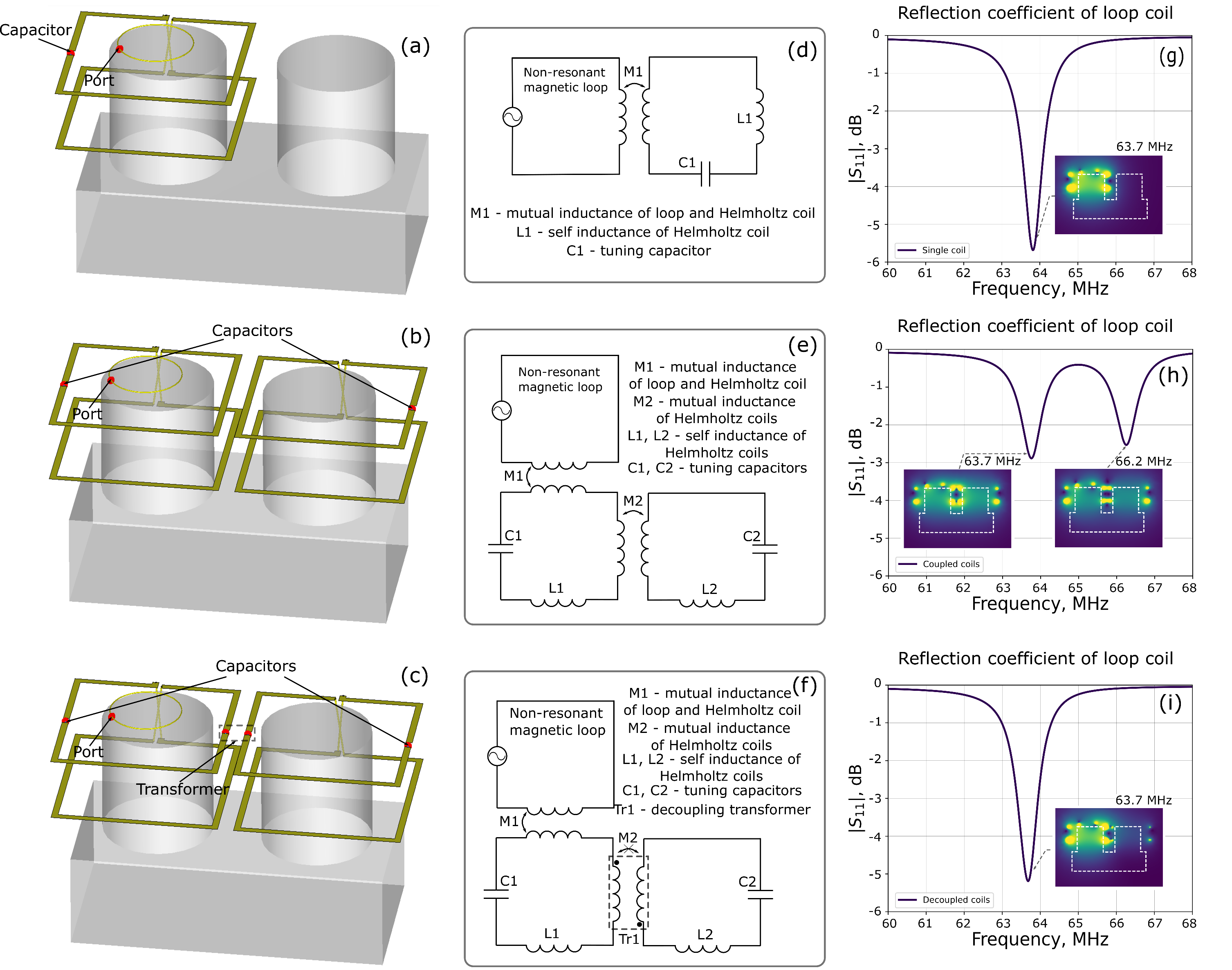}
\caption{Numerical simulations of wireless coils: a main view of a model containing homogeneous phantom and a single Helmhotlz coil (a), a pair of mutually coupled (b) and decoupled (c) Helmholtz coils inductively coupled with non-resonant coil; respective electrical functional circuits of single Helmholtz coil (d), a pair of mutually coupled (e) and decoupled (f) Helmholtz coils inductively coupled with non-resonant coil; reflection coefficient spectra of non-resonant loop inductively coupled with a single Helmholtz coil (g), a pair of mutually coupled (h) and decoupled (i) Helmholtz coils inductively coupled with non-resonant coil}
\label{Fig2}
\end{figure*}

A single Helmholtz resonator (coil) model loaded with a homogeneous phantom (Figure \ref{Fig2}b) was created as a reference setup. The dimensions of the Helmholtz coil loops are 152 x 157 mm, and the distance between them is 50 mm. To evaluate resonant characteristics and electromagnetic field distribution of a Helmholtz coil it was inductively coupled with a non-resonant feeding loop with with a 50 Ohm discrete port. The loop has a radius of 33 mm and is located at 17 mm above the phantom according to Figure \ref{Fig2}a. Lumped capacitor $C_{1}$ = 3.7 pF (Figure \ref{Fig2}d) is used to tune a Helmholtz coil to the resonant frequency of 63.7 MHz. The homogeneous phantom in simulations has an averaged material properties of a human breast with relative permeability of 40 and an electric conductivity of 0.3 S/m. 

The main view of the numerical model containing a pair of coupled Helmholtz coils is shown in Figure \ref{Fig2}b. Due to the close proximity of Helmholtz coils strong mutual coupling is expected. The respective electric functional schematic is shown in Figure \ref{Fig2}e, where mutual inductance between Helmholtz coils is denoted as $M_{2}$. To compensate mutual inductance of a pair of Helmholtz coils, a transformer decoupling was used (Figures \ref{Fig2}c,f). 

\begin{figure*}[ht]
\centering
\includegraphics[width=0.9\linewidth]{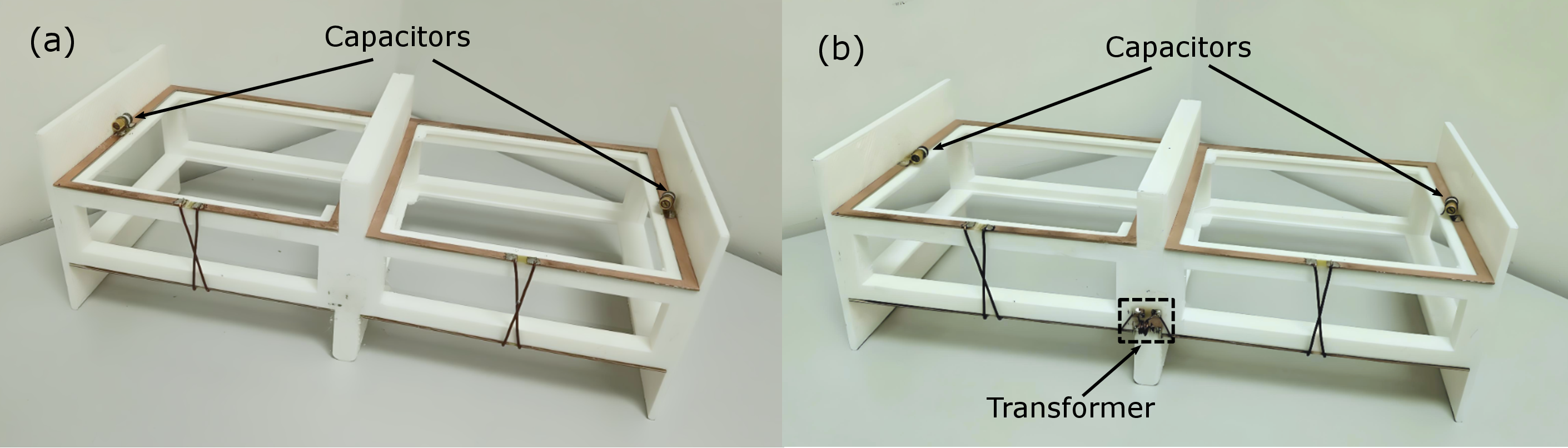}
\caption{Photo of assembled wireless coils prototypes: a pair of coupled (a) and decoupled (b) Helmholtz coils}
\label{Fig3}
\end{figure*}

To evaluate $|B_1^+|$ and local SAR distributions numerical simulations with a human voxel model Ella from the CST Voxel Family ~\cite{christ2009virtual} were performed. In these simulations, bilateral coil configurations inductively couple with a 600 mm long high-pass birdcage coil with 16 legs and a diameter of 700 mm. It was driven in quadrature by two 50-Ohm discrete ports embedded in parallel with two end-ring 43-pF tuning capacitors. The main goal of these simulations was to evaluate the influence of Helmholtz coil decoupling on their sensitivity and SAR values. 

\subsection{Prototypes and on-bench measurements}

To validate experimentally wireless coils and to make a comparison with simulations, two prototypes were assembled. The first one, contained two coupled Helmholtz coils is shown in Figure \ref{Fig3}a. The prototype is assembled using a 3D-printed frame made of a PLA plastic. Helmholtz coils loop conductors of 6 mm width were implemented using a FR-4 PCB 1-mm thick. Copper wires with 1.1 mm radius connect two loops of each Helmholtz resonator. To tune Helmholtz coils to the resonant frequency of 63.7 MHz trimmer (1.5-19 pF) non-magnetic capacitors (55H01, Knowles Johanson) were used. The second prototype consists of two Helmholtz coils inductively decoupled from each other using a transformer. To assemble inductors of a transformer, a copper wire with a radius of 0.9 mm was used. 

\begin{figure*}[ht]
\centering
\includegraphics[width=0.6\linewidth]{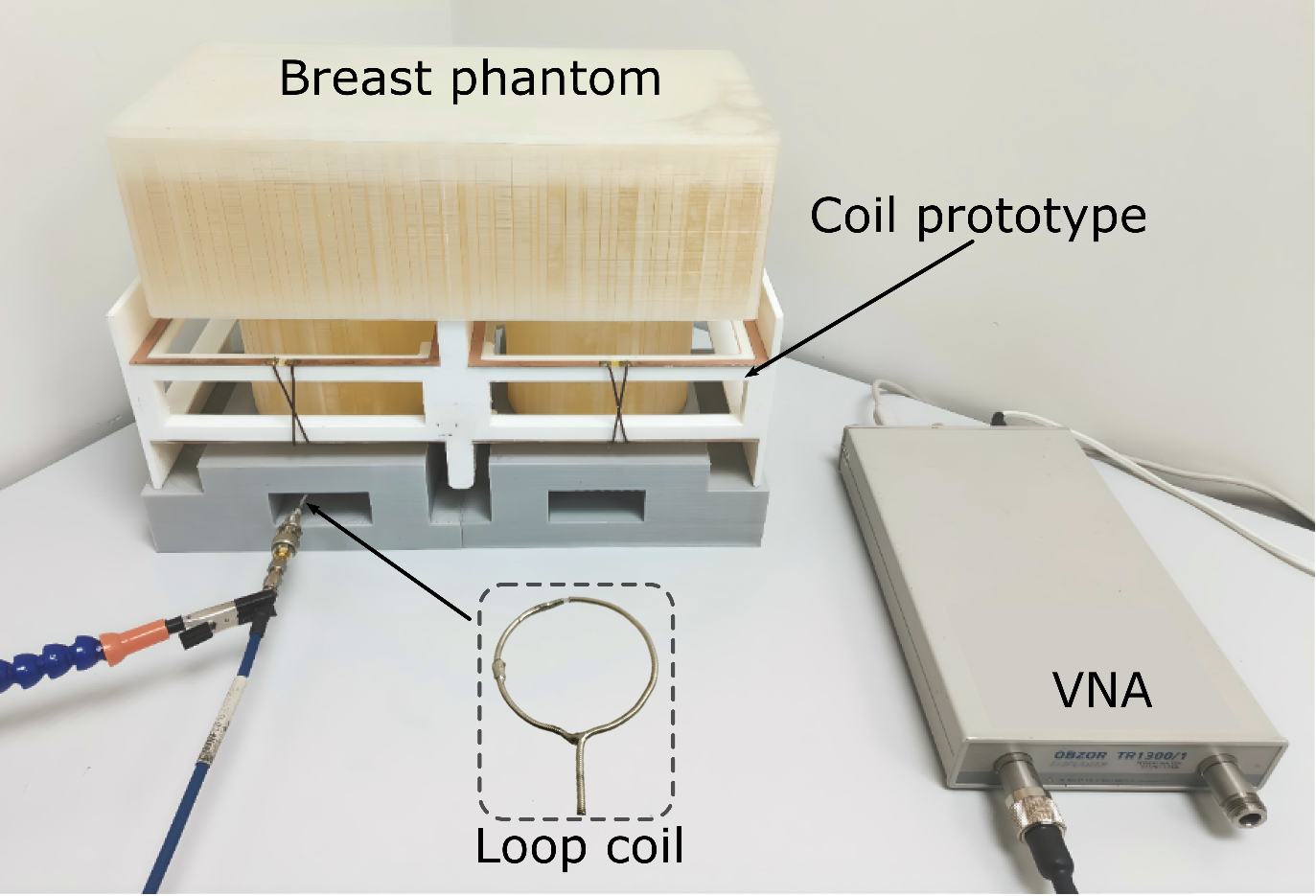}
\caption{Experimental setup for evaluation of resonant properties of wireless coils}
\label{Fig4}
\end{figure*}

Experimental validation of the resonant properties of wireless coils was performed using a setup shown in Figure \ref{Fig4}. The setup in the experiment is similar to numerical simulations. In particular, to provide a proper loading of wireless coils a homogeneous phantom was manufactured. A container for phantom liquid was assembled by a 3D-printing with PLA plastic. To obtain averaged material properties similar to the ones used in simulations (conductivity of 0.3 S/m and relative permittivity of 40), a solution containing 31.7\% distilled water, 4.8\% NaCl and 63.5\% sugar was manufactured.

To evaluate resonant conditions of coupled and decoupled Helmholtz coils non-resonant magnetic loop was used connected to the input of the vector network analyzer (VNA). In the experiment, the loop was placed under one of the Helmholtz coils pair according to Figure \ref{Fig4}. $S_{11}$ spectra near the working frequency of 63.8 MHz was evaluated using the VNA for a pair of coupled Helmholtz coils and a pair decoupled by a transformer. 

\subsection{MRI studies}

In the first round of MRI studies wireless coils were evaluated using a homogeneous phantom manufactured previously. These experiments were performed on a 1.5T Siemens MAGNETOM Espree (Erlangen, Germany). SNR and flip angle (FA) maps of homogeneous phantom were obtained for pairs of coupled and decoupled Helmholtz coils. FA maps were obtained by means of two gradient echo (GRE) acquisitions (TR/TE=2000/4.76~ms, FOV=${326}\times{326}$~mm$^{2}$, slice thickness=3~mm, acquisition matrix=${128}\times{128}$, $\text{FA}$=20$^{\circ}$ and $\text{FA}$=40$^{\circ}$) by the double-angle method on a pixel-by-pixel basis ~\cite{cunningham2006saturated}: 
\begin{equation}
FAm = \frac{arccos(\frac{S_{(FA=40^{\circ})}}{S_{(FA=20^{\circ})}})}{\pi}\cdot180^{\circ} ,
\end{equation}
where ${S_{FA=40^{\circ}}}$ is a magnitude of a pixel in MR image obtained with $FA=40^{\circ}$, and ${S_{FA=20^{\circ}}}$ - with $FA=20^{\circ}$; $FAm$ - is a measured point-by-point flip angle proportional to $|B_1^+|$ values. SNR maps were calculated using the MR images also on a pixel-wise basis where signal magnitude values were divided by the standard deviation of noise.



The second round of MR tests was performed in healthy volunteers using the same MR platform. Axial MR images of two volunteers were acquired using a decoupled wireless coil and a $T_{2}$-weighted turbo spin echo pulse sequence (SPAIR) with $\text{FA}$=160$^{\circ}$, TR/TE=6890/67~ms, FOV=${340}\times{340}$~mm$^{2}$, acquisition matrix=${384}\times{384}$, and slice thickness=4mm. 

\section{Results}
\subsection{Electromagnetic simulations}

The calculated reflection coefficient spectra of non-resonant loop inductively coupled with a single Helmholtz resonator is shown in Figure \ref{Fig2}g. When the second resonator is located electrically close to the first one, to make a bilateral coil, inductive coupling occurs illustrated by the reflection coefficient spectra shown in Figure \ref{Fig2}h. The resulting hybrid eigenmodes causes the occurrence of two resonance peaks at a loop coil $S_{11}$ spectra. Eigenfrequencies of these modes are determined by the values of tuning capacitors $C_{1}$ and $C_{2}$ of a Helmholtz pair (\ref{Fig2}e). These capacitors were tuned in order to adjust the low-frequency eigenmode (first mode) to 63.7 MHz. The resulting values of tuning capacitors are $C_{1}$=$C_{2}$= 3.22 pF. To study the effect of coil decoupling, a transformer was embedded between Helmholtz coils. Decoupling was observed using a transformer with inductances L = 103 nH and the mutual coupling coefficient equal to 0.4. The decoupling is observed as a degeneration of hybrid modes, resulting in the absence of resonance splitting in a probe loop $S_{11}$ spectra, as it is shown in Figure \ref{Fig2}i.

\begin{figure*}[ht]
\centering
\includegraphics[width=0.7\linewidth]{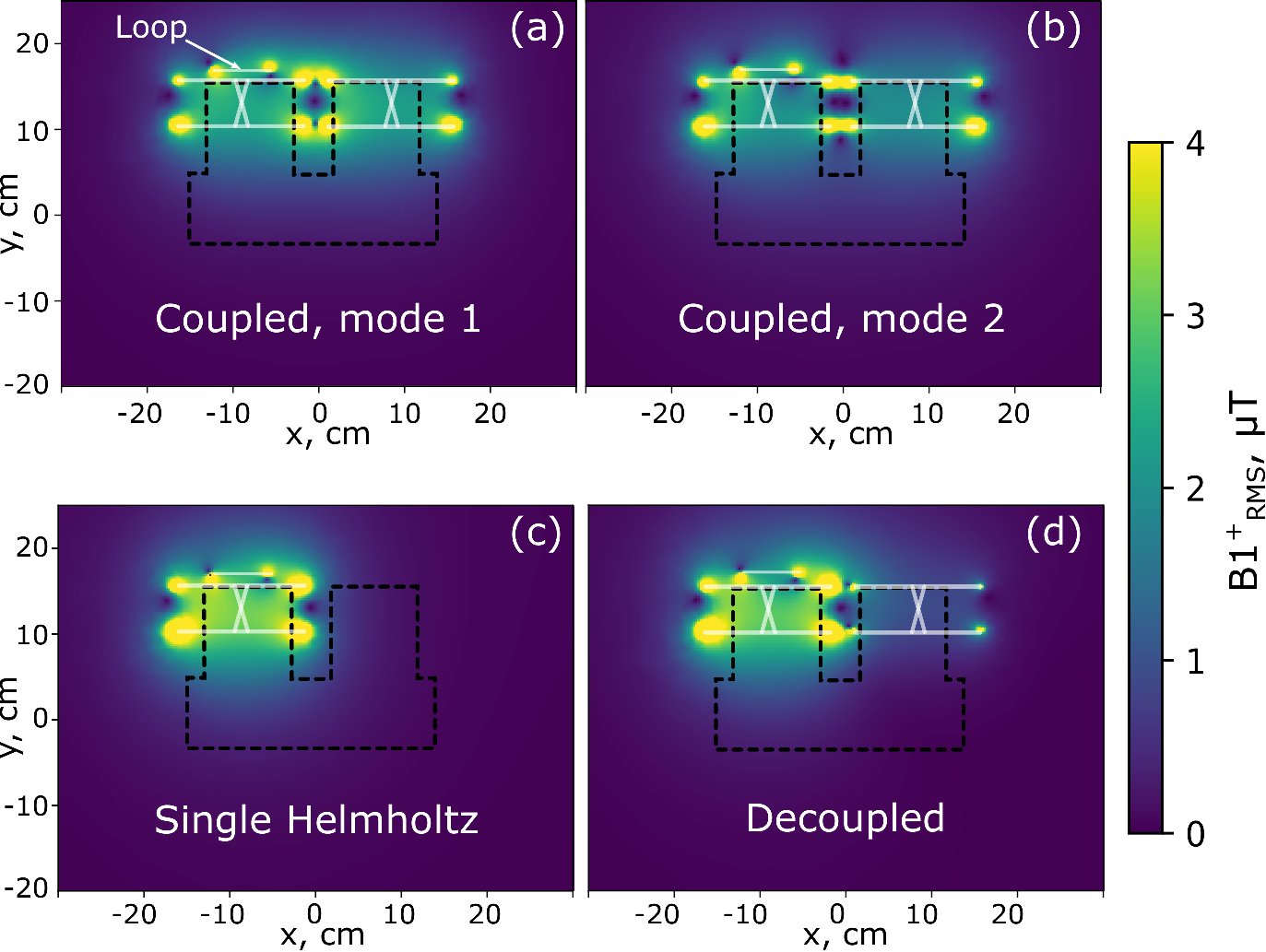}
\caption{Simulated $|B_1^+|$ distributions corresponding to the central transverse plane of a homogeneous phantom calculated for the pair of coupled Helmholtz coils plotted at the resonant frequency of the first (a) and second (b) hybrid mode, of a single Helmholtz coil (c) and of a pair of decoupled Helmholtz coils (d)}
\label{Fig5}
\end{figure*}

$|B_1^+|$ distributions were calculated for three setups, including single Helmholtz coil and pairs of coupled and decoupled coils. The resulting normalized to 1W of $P_{acc}$ $|B_1^+|$ distributions are summarized in Figure \ref{Fig5}. $|B_1^+|$ map corresponding to the first and second hybrid modes are shown in Figures \ref{Fig5}a and b, respectively. First mode is a symmetric one characterized by in-phase current distribution in two coils, while for the second antisymmetric mode, currents flow in opposite directions in these coils. Nevertheless, according to Figures \ref{Fig5}a and b, both modes are characterized by a homogeneous $|B_1^+|$ distribution in ROI. The difference occurs in ROI mean $|B_1^+|$ values, those are 2.47 uT for the first mode and 2.15 uT for the second mode.

\begin{figure*}[ht]
\centering
\includegraphics[width=1\linewidth]{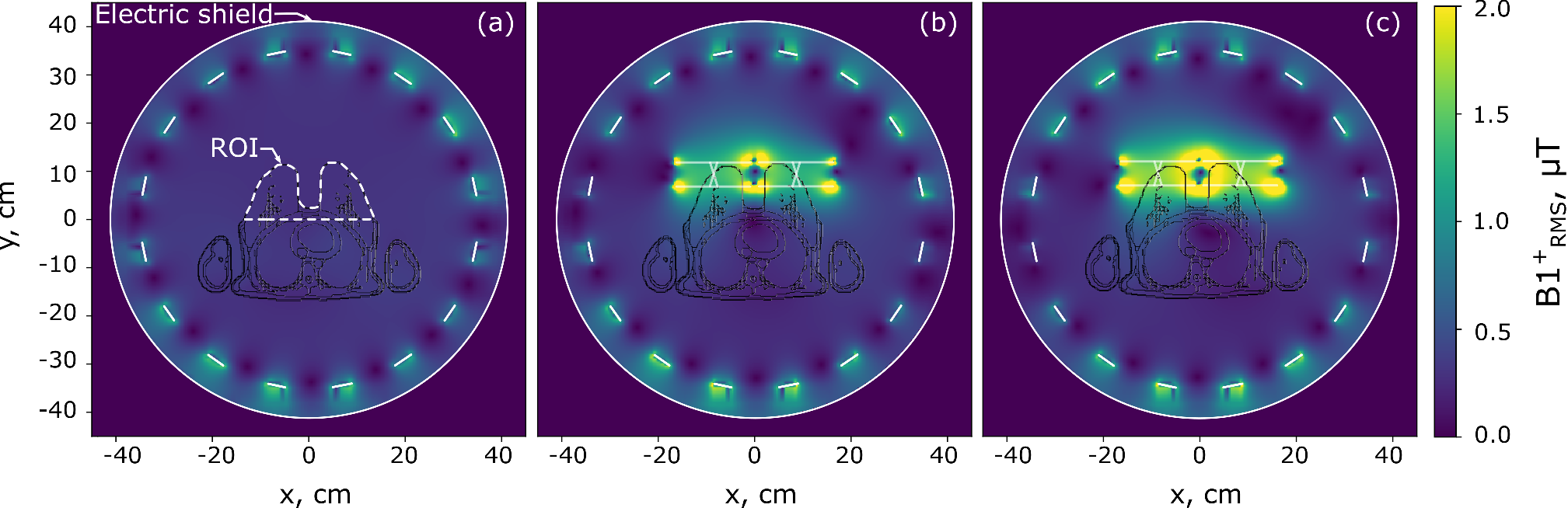}
\caption{Simulated $|B_1^+|$ distributions obtained using a voxel model corresponding to the central transverse plane of a breast calculated for BC only (a), for BC and a pair of coupled Helmholtz coils plotted at the resonant frequency of the first hybrid mode (b), for BC and a pair of decoupled Helmholtz coils (c)}
\label{Fig6}
\end{figure*}

$|B_1^+|$ distribution of the single Helmholtz coil is shown in Figure \ref{Fig5}c and the corresponding mean $|B_1^+|$ value in ROI is 3.31 uT. $|B_1^+|$ map of a pair of decoupled Helmholtz resonators is shown in Figure \ref{Fig5}d with 3.18 uT mean value in ROI which is 96\% of the reference single Helmholtz coil $|B_1^+|$ value. For the pair of coupled Helmholtz coils $|B_1^+|$ is 75\% and 65\%, compare to a single coil for the first and second mode, respectively.

\begin{figure*}[ht]
\centering
\includegraphics[width=1\linewidth]{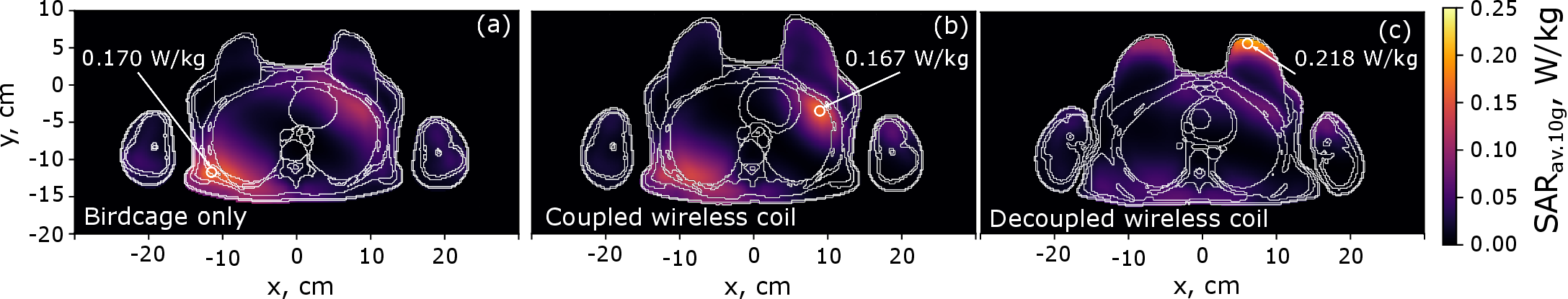}
\caption{Simulated local SAR$_{10g}$ distributions plotted in axial cross-section corresponding to the maximum SAR$_{10g}$ value in a voxel model for  BC only (a), for BC and a pair of coupled Helmholtz coils plotted at the resonant frequency of the first hybrid mode (b), for BC and a pair of decoupled Helmholtz coils (c)}
\label{Fig7}
\end{figure*}

The results of electromagnetic simulations using a voxel model are illustrated by the $|B_1^+|$ distributions shown in Figure \ref{Fig6} and the local SAR distributions shown in Figure \ref{Fig7}. The simulated $|B_1^+|$ map of a BC only, tuned to the resonant frequency of 63.7 MHz, is shown in Figure \ref{Fig6}a, illustrating the reference case with 0.3 uT mean value in the target ROI. This value is obtained when the conventinal clinical setup is used with BC used for transmission and Rx-only coil for the reception of MR signal. The $|B_1^+|$ distribution calculated for a pair of coupled Helmholtz coils tuned to the first hybrid mode excited by a BC is shown in Figure \ref{Fig6}b, and for a pair of decoupled Helmholtz coils in Figure \ref{Fig6}c. The mean $|B_1^+|$ values in ROI are 1.04 uT and 1.29 uT for the coupled and decoupled Helmholtz coils, respectively. The applied decoupling strategy allows the mean $|B_1^+|$ value improvement in ROI at 24\%, confirming results, obtained previously using simulations with homogeneous phantom. 

Local $SAR_{10g}$ distributions, corresponding to local maximum planes, normalized to 1 W of Pacc are shown in Figure \ref{Fig7}. Reference local $SAR_{10g}$ distribution of BC only is depicted in Figure \ref{Fig7}a, while for the pairs of coupled and decoupled resonators in Figures \ref{Fig7} b and c correspondingly. The maximum local $SAR_{10g}$ value for BC only is observed in a spine area and is equal to 0.170. For the pair of coupled Helmholtz coils the respective value is 0.167 W/kg occurring in a chest, while for the pair of coupled resonators it is 0.218 W/kg in the area of breast. All the obtained local $SAR_{10g}$ values are well below the allowed limits, showing that the safety issues are not compromised. The results of numerical simulation are summarized in Table \ref{Tab_1}. 

\begin{table}[]
\centering{}
\begin{tabular}{|c|c|cc|}
\hline
\multirow{2}{*}{}       & phantom with loop & \multicolumn{2}{c|}{voxel model with BC}                      \\ \cline{2-4} 
                        & $|B_1^+|$, uT     & \multicolumn{1}{c|}{$|B_1^+|$, uT} & SAR$_{\text{10g}}$, W/kg \\ \hline
BC only                 & -                 & \multicolumn{1}{c|}{0.3}           & 0.170                    \\ \hline
Single coil             & 3.31 (100\%)      & \multicolumn{1}{c|}{-}             & -                        \\ \hline
Decoupled coil          & 3.18  (96\%)      & \multicolumn{1}{c|}{1.29}          & 0.218                    \\ \hline
Coupled coil (1st mode) & 2.47 (75\%)       & \multicolumn{1}{c|}{1.04}          & 0.167                    \\ \hline
Coupled coil (2nd mode) & 2.15 (65\%)       & \multicolumn{1}{c|}{-}             & -                        \\ \hline
\end{tabular}
\caption{Results of numerical simulations  with $|B_1^+|$ and maximum local $SAR_{10g}$ values for different setups and coil configurations.}
\label{Tab_1}
\end{table}

\begin{figure*}[ht]
\centering
\includegraphics[width=0.9\linewidth]{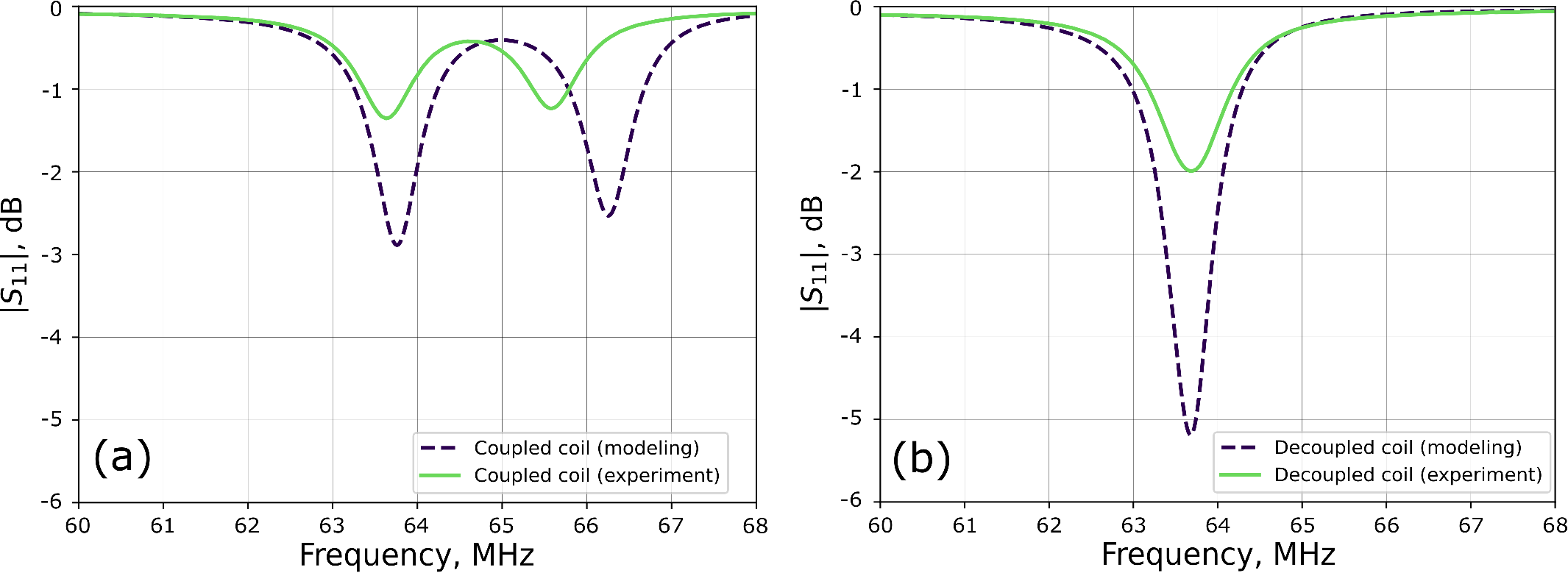}
\caption{Reflection coefficient $S_{11}$ spectra of the non-resonant magnetic loop inductively coupled with the one of the paired coupled Helmholtz coils (a) and he one of the paired decoupled Helmholtz coils (b), where dashed dark blue line corresponds to simulations, green solid line -- to an experiment}
\label{Fig8}
\end{figure*}

\subsection{Experimental evaluations}
\subsubsection{On-bench measurements}

The simulated pairs of coupled and decoupled pairs of Helmholtz coils were implemented in prototypes shown in Figure \ref{Fig3}. Using the experimental setup shown in Figure \ref{Fig4} resonant characteristics of Helmholtz coils pairs were evaluated. The obtained experimental $S_{11}$ curves of the loop coil inductively coupled with the one Helmholtz coil of each pair are shown in Figure \ref{Fig8} by dashed dark blue lines together with the simulated curves (solid green lines). In Figure \ref{Fig8}a $S_{11}$ spectra obtained for a pair of coupled Helmholtz coils, while in Figure \ref{Fig8}b -- for a pair of decoupled coils. The same hybridization of eigenmodes is observed for a pair of coupled Helmholtz resonators as it was predicted by numerical simulations. The differences between simulated and experimental $S_{11}$ values at resonant frequencies is most probably due to the different size of loop probes and slightly different distance between the loop probe and Helmholtz coil. The mismatch between hybrid eigenmodes frequencies in simulation and experiments is most likely due to the slight difference between the respective position of Helmholtz resonators in simulations and experiments as well as moderate deviation of coil loading.  


\subsubsection{MRI trials}

Both pairs of coupled and decoupled Helmholtz coils were studied in MRI using homogeneous phantom. Since wireless coils increase the transmit efficiency of a BC, the reference voltage in the inputs of a BC should be adjusted. The reference voltage for BC only was 318 V, while for pairs of coupled and decoupled Helmholtz coils it was 111.5 and 77.4 V, respectively.

\begin{figure*}[ht]
\centering
\includegraphics[width=0.7\linewidth]{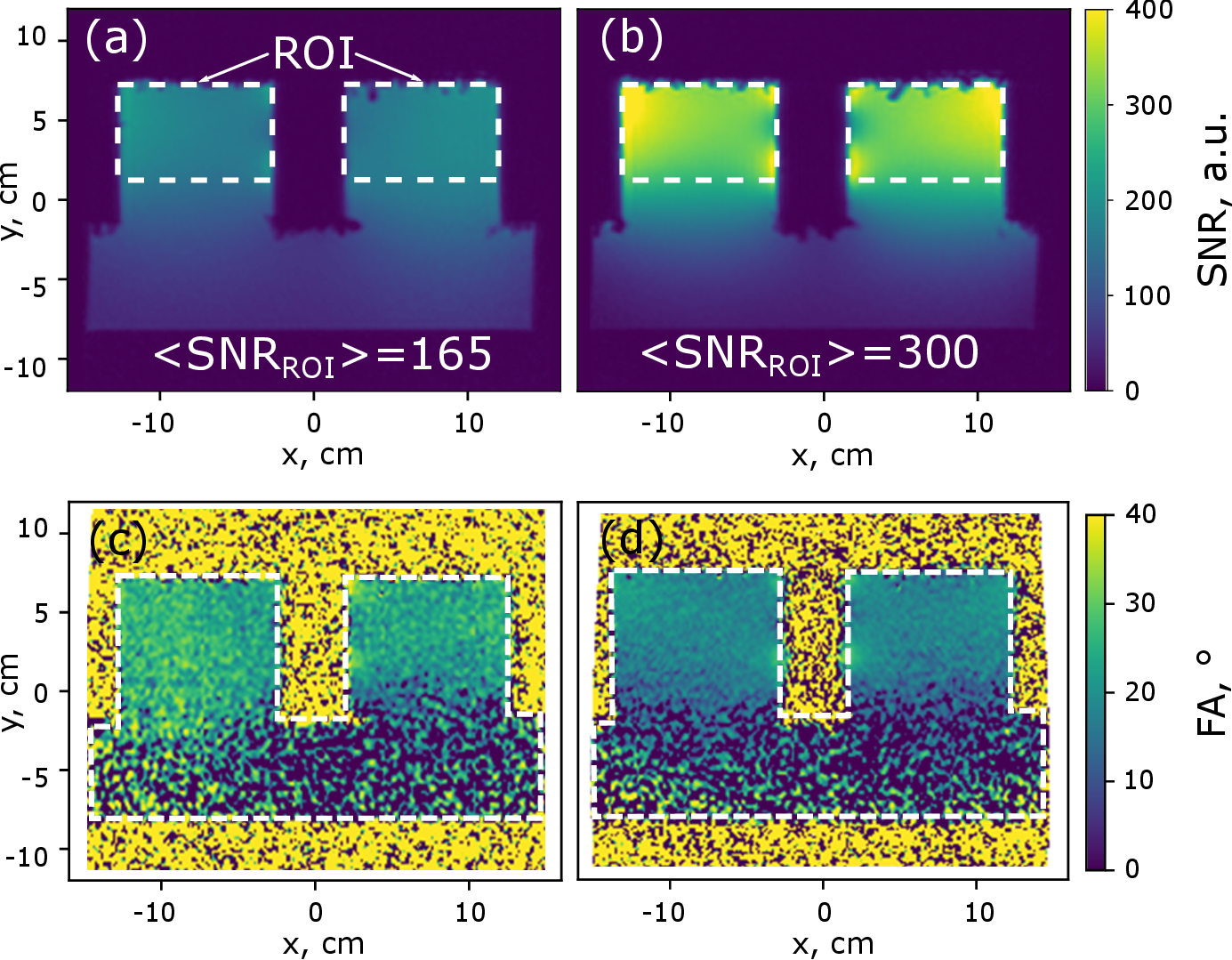}
\caption{Results of phantom imaging: SNR maps obtained for a pair of coupled (a) and decoupled Helmholtz coils (b); FA maps calculated by the double-angle method for a pair of coupled (c) and decoupled Helmholtz coils (d)}
\label{Fig9}
\end{figure*}

Figure \ref{Fig9} shows SNR maps, obtained for both coil configurations by a GRE sequence for $FA=90^{\circ}$: for the pair of coupled (Figure \ref{Fig9}a) and for the pair of decoupled coils (Figure \ref{Fig9}b). The average SNR in ROI for coupled coil and decoupled coil is equal to 165 and 300 respectively. The obtained using a GRE sequence and a double angle method FA maps ($FA=40^{\circ}$ and $FA=20^{\circ}$) are shown in Figure \ref{Fig9}. Namely, in Figure \ref{Fig9}c FA map for the pair of coupled coils and in Figure \ref{Fig9}d for the pair of decoupled coils are shown. The average FA value in the left half of the coupled pair is 25 $\pm$ 3.8, in the right half is 23 $\pm$ 3.6, while for the decoupled pair these values are 18.7 $\pm$ 2.1 and 18.4 $\pm$ 2.2 for the left and right halves, respectively. The target FA value in this study is equal to 20, and for the pair of decoupled coils the obtained values are closer to the expected values. Moreover, due to the results obtained, $|B_1^+|$ is more homogeneous for a pair of coupled coils.

\begin{figure*}[ht]
\centering
\includegraphics[width=1\linewidth]{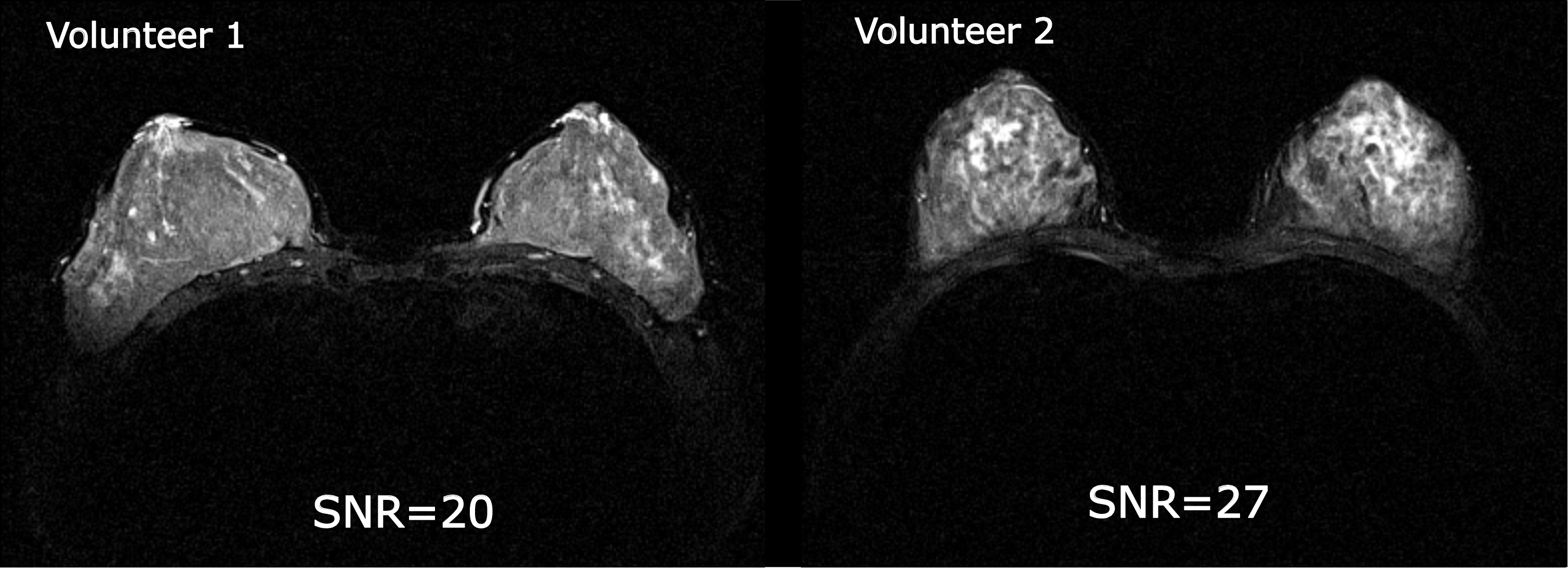}
\caption{In vivo  MR images obtained for two healthy volunteers by a pair of decoupled Helmholtz coils and a $T_{2}$-weighted turbo spin echo pulse sequence SPAIR}
\label{Fig10}
\end{figure*}

To test the pair of decoupled coils using the conditions of a routine MRI procedure, in vivo studies with two healthy volunteers were performed. The reference voltage for the first volunteer was 123 V with a pair of decoupled Helmholtz coils and 373 V for BC only, the reference voltage for the second volunteer was 107 V with a pair of decoupled coils and 374 V for BC only. In vivo MR images obtained using a  $T_{2}$-weighted turbo spin echo pulse sequence SPAIR are shown in Figure \ref{Fig10}. For the first volunteer, the obtained SNR in ROI is 20, while for the second volunteer it is equal 27. The results of experimental evaluations are summarized in Table \ref{Tab_2}.   

\begin{table}[]
\centering
\resizebox{\textwidth}{!}{%
\begin{tabular}{|c|ccc|ccc|cc|}
\hline
\multirow{3}{*}{} & \multicolumn{3}{c|}{Reference voltage}                                         & \multicolumn{3}{c|}{SNR}                                                     & \multicolumn{2}{c|}{FA}                    \\ \cline{2-9} 
                  & \multicolumn{2}{c|}{volunteer}                      & {phantom} & \multicolumn{2}{c|}{volunteer}                    & {phantom} & \multicolumn{2}{c|}{phantom}               \\ \cline{2-3} \cline{5-6} \cline{8-9}  
                  & \multicolumn{1}{c|}{1st}   & \multicolumn{1}{c|}{2nd}   &         & \multicolumn{1}{c|}{1st}  & \multicolumn{1}{c|}{2nd}  &         & \multicolumn{1}{c|}{Left}         & Right         \\ \hline
BC only           & \multicolumn{1}{c|}{373} & \multicolumn{1}{c|}{374} & 319     & \multicolumn{1}{c|}{-}  & \multicolumn{1}{c|}{-}  & -                        & \multicolumn{1}{c|}{-}         & -         \\ \hline
decoupled coil    & \multicolumn{1}{c|}{123} & \multicolumn{1}{c|}{107} & 77.4    & \multicolumn{1}{c|}{20} & \multicolumn{1}{c|}{27} & 300     & \multicolumn{1}{c|}{18.7 $\pm$ 2.2} & 18.4 $\pm$ 2.2 \\ \hline
coupled coil      & \multicolumn{1}{c|}{-}   & \multicolumn{1}{c|}{-}   & 111.5                    & \multicolumn{1}{c|}{-}  & \multicolumn{1}{c|}{-}  & 165                      & \multicolumn{1}{c|}{25.0 $\pm$ 3.8} & 23.0 $\pm$ 3.6 \\ \hline
\end{tabular}%
}
\caption{Resulting characteristics designed coils in MRI experiment.}   
\label{Tab_2}
\end{table}




\section{Discussion and conclusion}
In this work, we study the influence of a pair of Helmholtz coils decoupling on the performance of bilateral wireless coil. We have shown numerically and experimentally an improvement of a transmit efficiency and receive sensitivity of decoupled Helmholtz coils compare to coupled ones. The results of electromagnetic simulations with phantom shows the positive impact of Helmholtz coils decoupling on the performance of thereof. An average $|B_1^+|$ of decoupled Helmholtz coils is 4 \% smaller than average $|B_1^+|$ of a single Helmholtz coil, while an average $|B_1^+|$ for the coupled coil when the first and second mode are excited is 25 \% and 35 \% respectively smaller compare to a single Helmholtz coil. Current flowing in a Helmholtz coil of a coupled pair causes the magnetic field in the area of a second Helmholtz coil. The resulting inductive coupling causes a current flowing in the second coil in the opposite direction than the source current that generated it. The same effect is observed in the second coil of a pair with respect to the first one. The resulting secondary currents cause out-of-phase magnetic fields with respect to the primary magnetic fields in the area of each resonator of a coupled pair. The resulting destructive interference of primary and secondary magnetic fields reduces the resulting $|B_1^+|$ of a coupled wireless coil.

This is confirmed also by the results of electromagnetic simulations with a voxel model and wireless coils inductively coupled with BC, showing that for a pair of decoupled Helmholtz coils $|B_1^+|$ in ROI is 24 \% higher compare to a pair of coupled Helmholtz coils. In terms of local SAR values, the coupled Helmholtz pair shows 0.167 W/kg local maximum, while a pair of decoupled Helmholtz coils demonstrates 0.218 W/kg. Hence the maximum local SAR value is higher for the pair of coupled coils, both values are well below the standard limits. 

The results of on-bench investigations show that the wireless coil is sensitive to the assembly accuracy, illustrated by the differences of a resonant frequency split for the pair of coupled coils prototype compare to numerical simulations. 


The results of MRI trials with phantom show, that average SNR in ROI for a pair of decoupled coils is 82 \% higher than for a pair of coupled coils. The obtained FA map for a pair of the decoupled coils demonstrate higher homogeneity (left half $|B_1^+|$ StD = 2.2${^\circ}$, right half $|B_1^+|$ StD = 2.2${^\circ}$) and average FA in ROI (left half mean FA = 18.7${^\circ}$, right half mean FA = 18.4${^\circ}$) closer to a desired FA value = 20${^\circ}$, compare to a pair of the decoupled coils (left half $|B_1^+|$ StD = 3.8${^\circ}$, right half $|B_1^+|$ StD = 3.6${^\circ}$; left half mean FA = 25${^\circ}$, right half mean FA = 23${^\circ}$). The experimental SNR ratio between the pair of decoupled coils and coupled ones is higher compare to numerically obtained $|B_1^+|$ ratio between these coils. The expected increase of average SNR value for decoupled coils is approximately 20 \% -- 25 \% compare to coupled coils. The difference in reference voltage is 44 \%, which better matches with the results of numerical simulations. We suppose, that the mismatch in SNR ratio between experiment and simulations is caused by suboptimal mode tuning for a pair of coupled Helmholtz coils in MRI experiment.
The clinically relevant images were also obtained for two healthy volunteers using decoupled Helmholtz coils demonstrating promising solution for breast imaging applications.

\section*{Acknowledgements}
This work was supported by state assignment No. FSER-2022–0010 within the framework of the national project “Science and Universities”.

\section*{Conflict of interest}
The authors declare no conflict of interest.

\bibliography{main}




\end{document}